# Spatial resolution in optical coherence elastography of bounded media

GABRIEL REGNAULT,[1] MITCHELL A. KIRBY,[1*] MAJU KURIAKOSE,[1] TUENG T. SHEN,[2] RUIKANG K. WANG,[1,2] MATTHEW O'DONNELL,[1] IVAN PELIVANOV[1]

[1]*University of Washington, Department of Bioengineering, Seattle, Washington, 3720 15th Ave NE, Seattle, WA 98105, USA*
[2]*University of Washington, Department of Ophthalmology, Seattle, Washington, 1959 NE Pacific St, Seattle, WA 98195 USA*
**makirby@uw.edu*

**Abstract:** Dynamic optical coherence elastography (OCE) tracks mechanical wave propagation in the subsurface region of tissue to image its shear modulus. For bulk shear waves, the lateral resolution of the reconstructed modulus map (i.e., elastographic resolution) can approach optical coherence tomography (OCT) capabilities, typically a few tens of microns. Here we perform comprehensive numerical simulations and acoustic micro-tapping OCE experiments to show that for the typical situation of guided wave propagation in bounded media, such as cornea, the elastographic resolution cannot reach the OCT resolution and is mainly defined by the thickness of the bounded tissue layer. We considered the excitation of both broadband and quasi-harmonic guided waves in a bounded, isotropic medium. Leveraging the properties of broadband pulses, a robust method for modulus reconstruction with minimum artifacts at interfaces is demonstrated. In contrast, tissue bounding creates large instabilities in the phase of harmonic waves, leading to serious artifacts in modulus reconstructions.

## 1. Introduction

The speed of a propagating mechanical wave in a medium is fully defined by its elastic modulus and the boundary conditions. Often, boundary conditions are ignored (i.e., bulk plane-wave propagation in an infinite medium is assumed) in dynamic elastography such that the mechanical modulus at every sample point along a propagation path can be fully defined by the measured local wave speed. If the assumption of plane wave propagation in a bulk medium is accurate, spatially resolved mechanical modulus reconstruction is a simple one-to-one mapping from the locally measured wave speed. Indeed, many soft biological tissues are nearly incompressible and are characterized by a single mechanical modulus, which can be computed from the measured shear wave speed through a simple expression [1]–[3].

For the ideal case where plane shear waves are used and interface waves can be ignored, the spatial resolution of the reconstructed modulus is limited by the capabilities of the imaging system. For example, spatial resolution in conventional ultrasound elastography (USE) is primarily determined by the scanning beam width (on the order of a few hundreds of micrometers) [4], while optical coherence tomography (OCT) has a lateral resolution determined by the optical focusing power of the scanning lens and can be on the order of microns [5]. OCT-based elastography (OCE) holds the promise for high-resolution modulus reconstruction due to its high-resolution imaging, non-contact nature, and ability to detect nanometer scale vibrations [1], [6]–[11]. Compared to USE, OCE has the potential to improve elastographic resolution by an order of magnitude. Measured resolution in wave-based OCE has never approached the ultimate resolution of the OCT system, however, even though OCE resolution is often assumed to equal that of the OCT imaging system.

In this work, we define elastographic resolution as the minimum inclusion size for which the mechanical modulus can be accurately reconstructed with little or no a-prior knowledge of material stiffness. OCE usually images shallow subsurface tissue regions and, therefore, probes surface propagating mechanical waves. As shown in Ref. [1], the mechanical resolution in dynamic OCE utilizing surface (or Rayleigh) mechanical waves is generally limited by the characteristics of the mechanical wave (frequency, bandwidth) and not by the capabilities of the OCT system (assuming sufficient temporal and spatial sampling for OCT reconstruction of light scattering and wavefront propagation). In heterogeneous materials, Rayleigh waves mix with different kinds of waves generated at boundaries, for example at the edges of an inclusion, that complicates wave analysis. Mode conversions and guided propagation can lead to serious errors in the reconstructed elastic modulus [1], [2]. As shown previously, the lateral resolution in elastic modulus maps reconstructed from surface wave propagation captured with dynamic OCE is about half of the mechanical wavelength, or on the order of the mechanical pulse width (for broadband signals) [1] .

Because OCT is generally restricted to surface measurements (on the order of 2-3 mm in depth), and many tissues have a layered structure, wave-based OCE methods do not exclusively use Rayleigh waves, i.e. waves propagating over the surface of a bulk material. Many tissues and organs (skin, cornea, gastro-intestinal channels, etc.) are bounded, leading to guided mechanical waves [3]–[5]. Guided waves are aggregate waves formed by reflections and reverberations of bulk mechanical waves inside a bounded layer. As shown previously, guided elastic waves are dispersive and require an appropriate mechanical model to reconstruct elastic moduli from them [2], [6].

In this study, we conducted numerical simulations in OnScale (Redwood City, California) and performed acoustic micro-tapping (A$\mu$T) OCE measurements to explore the spatial resolution of reconstructed modulus maps for a layered, isotropic medium where the thickness is on the order of the mechanical wavelength (i.e., guided wave propagation). The lateral resolution in quantitative modulus images is limited by the wavelength of the elastic wave, similar to the Rayleigh wave case, but additionally depends on the layer thickness.

## 2. Materials and Methods

### 2.1. Numerical Simulations

A finite element (FEM) simulation was designed to simulate ideal, noise-free guided wave behavior in a two-part isotropic medium (Fig.1). The simulation geometry is shown in Fig.1a. The lengths of the first and second parts are 10 mm and 20 mm, respectively. The shear modulus of part I was held constant at $G_1 = 30$ kPa, and the shear modulus of part II was set to different values of $G_2 = [33, 45, 60, 120]$ kPA for different simulations. The material density ($\rho = 1000$ kg/m$^3$) was uniform throughout both parts. Longitudinal wave speed was uniform ($C_l$ = 1540 m/s) throughout the entire domain (including water) except for air. Four different thicknesses $h = [0.125, 0.25, 0.5, 1]$ mm were investigated.

The bottom and right boundaries were set to be absorbing layers to minimize reflections. The domain was discretized with linear finite elements on a regular rectangular grid with a grid spacing of 5 $\mu$m for $h = [0.25, 0.5, 1]$ mm and 2.5 $\mu$m for $h = 0.125$ mm. Because under-integrated linear elements are prone to spurious solutions, we applied Belytchko-Bindeman strain hourglass suppression.[7] The equations were integrated in time using an explicit time-stepping method to generate the full vibration velocity vector (time derivative of the dynamic displacement vector) at each space and time point. Only the vertical component of the vibration velocity was analyzed for direct comparison to experimental results (see also Ref [8], [9] for details on setting simulations in OnScale).

Along the top boundary of the domain ($z = 0$), we applied a spatially- and temporally-varying pressure load to simulate the acoustic push applied to the surface by the A$\mu$T transducer in our experiments [1]. A line source was considered to mimic experimental push conditions [9]–[11]. The acoustic intensity $I(x)$ of the push across the line source was considered Gaussian with a full-width-at-half max (FWHM) of 500 μm. We used this Gaussian model to define the spatial push profile:

$$P(x) = \frac{(1+R^2)I(x)}{C_{air}} \approx \frac{2}{C_{air}} \cdot P_0 \cdot \exp\left[-4 \cdot \log 2 \cdot \left(\frac{x}{d}\right)^2\right], \quad (1)$$

where $R$ and $C_{air}$ are the reflection coefficient and $l$- wave speed of air, respectively ($R \approx 1$, $C_{air}$ = 340 m/s), $P_0$ = 5 kPa is the amplitude of the applied pressure, and $d$ = 500 μm is the push width, defined here as the FWHM of the acoustic intensity.

We simulated two different temporal excitation profiles. First, the temporal push profile was very short, i.e. shorter than the time required for a shear wave to propagate across the excitation zone. This satisfied the pressure confinement condition for excitation and launched ultra-broadband pulses of mechanical waves. In the second situation, the envelope of the temporal push was modulated by a harmonic law with a 1 kHz repetition rate.

### 2.1.1. Broadband excitation

For the broadband excitation, the push temporal profile followed a super-Gaussian function

$$s(t) = \exp\left(-16 \cdot \log 2 \cdot \left(\frac{t-t_0}{T}\right)^4\right), \quad (2)$$

where $T$ was the push duration, defined as the full-width-at-half-maximum of the envelope function $s(t)$. A time delay ($t_0$) was introduced to avoid impulsive loading at time $t = 0$. The push profile was centered, and full symmetry was assumed at the left boundary. To fulfill the pressure confinement condition for mechanical wave excitation, the duration of the push was considered short and equal to $T = 100 \ \mu s$.

Wave behavior in bounded media can be complicated and result in multiple guided modes (Fig.1b). Using the group velocity of propagating waves to evaluate the shear modulus can produce serious mistakes [2]. To accurately reconstruct the shear modulus from guided wave propagation, guided modes must be sorted. A common method for mode sorting computes the two-dimensional (2D) Fourier Transform of the space-time profile (or $XT$ plot, see Fig.1b) to obtain its wavenumber-frequency spectrum or ($kf$ plot, see Fig.1d). In the case of guided broadband propagation, the $kf$ plot clearly shows the two first modes of guided behavior ($A_0$ and $S_0$ modes).

The spatial range used in the computation of the $kf$ spectrum (shown as a window the in dashed rectangle in Figs.1b, e, h) defines the lateral spatial resolution in the reconstructed modulus. We used rectangular windows smoothed with a Gaussian function at both edges:

$$W(x,t) = w_x \cdot w_t, \quad (3)$$

where

$$w_x = \begin{cases} 1 \text{ if } X_{min} < x < X_{max} \\ \exp(-\frac{1}{2}\left(4\frac{x-X_{max}}{\sigma_x}\right)^2) \text{ if } x > X_{max} \\ \exp(\frac{1}{2}\left(4\frac{x-X_{min}}{\sigma_x}\right)^2) \text{ if } x < X_{min} \\ X_{min} = X_{center} - \frac{L_{win}}{2} + \sigma_x \\ X_{max} = X_{center} + \frac{L_{win}}{2} - \sigma_x \\ \sigma_x = \frac{L_{win}}{10} \end{cases} \text{ and } w_t = \begin{cases} 1 \text{ if } T_{min} < t < T_{max} \\ \exp(-\frac{1}{2}\left(4\frac{t-T_{max}}{\sigma_t}\right)^2) \text{ if } t > T_{max} \\ \exp(\frac{1}{2}\left(4\frac{t-T_{min}}{\sigma_t}\right)^2) \text{ if } t > T_{min} \\ T_{max} = L_t - \sigma_t \\ T_{min} = \sigma_t \\ \sigma_t = \frac{L_t}{10} \end{cases}$$

with $L_t$ the total recording time, and $L_{win}$ and $X_{center}$, respectively, the size and position of the window in the x-dimension.

Assuming appropriate temporal sampling of the propagating elastic wave, a relatively large window in space results in a localized $kf$ spectrum where guided wave modes are clearly defined. When guided modes are sampled sufficiently, reconstructed shear moduli are accurate. To demonstrate the influence of the spatial window on modulus reconstruction, we computed the energy in the spectrum along theoretical solutions for $A_0$ and $S_0$, for a broad range of shear modulus, $G$, values (this is later referred to as the goodness of fit, or GOF). The best estimate of the modulus, labelled $G_1$, corresponds to the maximum GOF. Figures 1c, f, i show how resolving guided modes is less accurate as the window size shrinks, resulting in a broadened spectra and corresponding GOF changes (Figs. 1d, g, j). When the size of the window used to reconstruct the modulus is smaller than a certain value, fitting is incorrect (Fig. 1i), and reconstruction of the shear modulus becomes impossible (Fig. 1j).

Note that only the first two guided modes were used for fitting. Although an infinite number of modes can be obtained in simulations for an infinitely short in time and infinitely narrow in space excitation, localization of the push in time and space is usually limited in practice. This in turn limits the bandwidth of launched mechanical transient and, therefore, the number of generated guided modes. Practical implementations of OCE are usually limited to $A_0$ and $S_0$ modes.[12]

The goal of numerical simulations was to define the minimum window size for which the computed $kf$ spectrum can still be fitted with a theoretical solution for $A_0$ and $S_0$ modes to produce accurate values of reconstructed shear modulus. We investigated how the minimum window size depends on the A$\mu$T push characteristics and the medium thickness. When the window size was determined, we 'moved' the window along the medium surface through the interface separating two distinct isotropic media (i.e. with different stiffnesses) to finally define the lateral resolution in OCE of bounded structures.

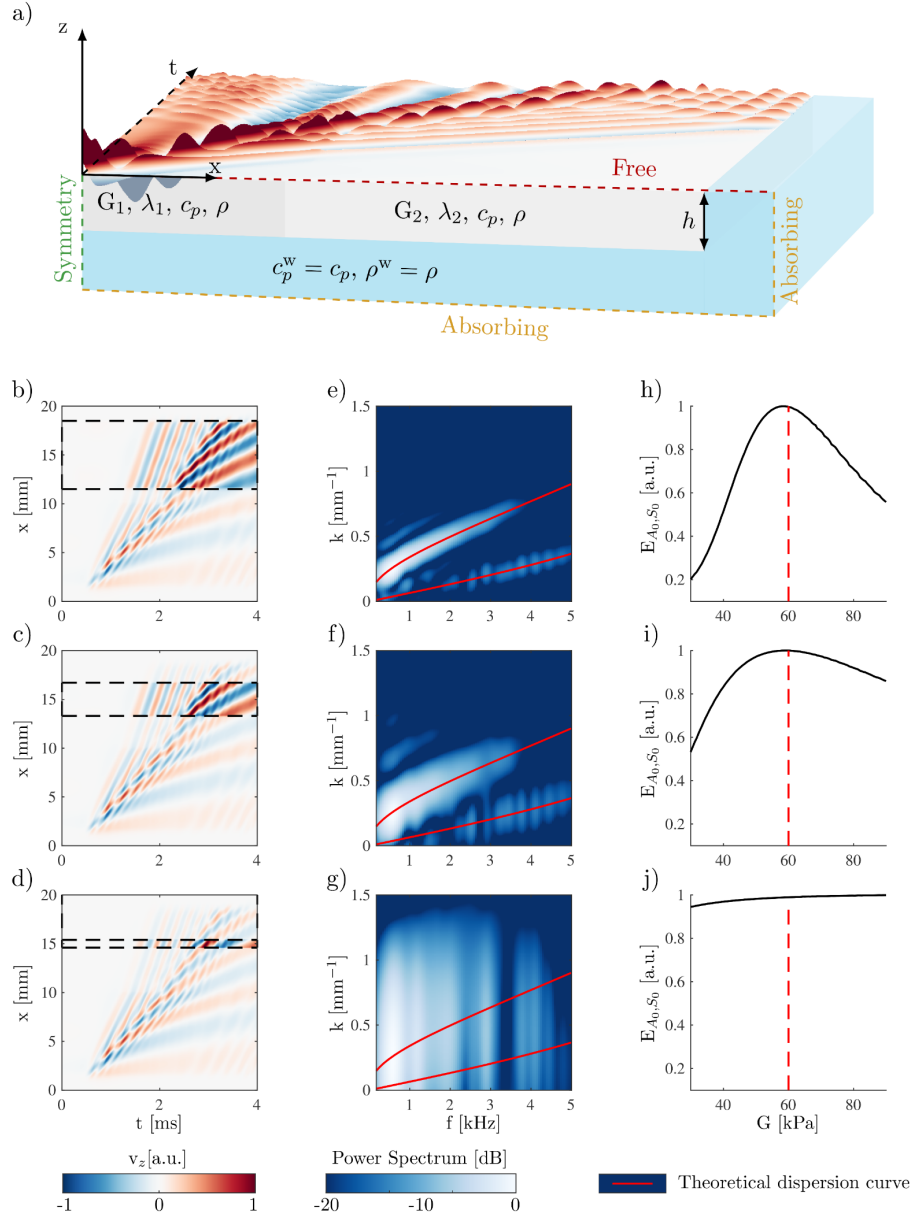

Figure 1. Numerical simulation in OnScale of guided wave propagation in a two-part nearly incompressible linear isotropic material of uniform thickness ($h = 0.5$ mm in this figure) bounded below by water and above by air (mimicking cornea boundary conditions). Shear moduli in parts I and II used in this figure are equal to $G_1 = 30$ kPa and $G_2 = 60$ kPa respectively; medium thickness is uniform over the medium and equal to $h = 0.5$ mm. The interface between the two parts is at $X = 10$ mm. Broadband pulses of guided waves were launched with acoustic micro-tapping [10] from air. The push profile was considered Gaussian in space and super-Gaussian in time, as described in Eq. (1). (a) Medium geometry used in the OnScale simulations. A 1-mm layer of water bounds the media on the right-hand side to allow a better match in the absorbing boundary conditions. (b) Space-time ($XT$) plot of guided wave propagation at the medium surface. (c) Wavenumber-frequency ($kf$) spectrum obtained from $XT$ plot (b) using an 8 mm window in part II with solutions for $A_0$ and $S_0$ modes best fitting the spectrum. (d) Goodness of fitting (GOF) of the simulated spectrum as a function of $G_1$ describing the method of shear modulus reconstruction from the $kf$ spectrum. (e), (f), (g) are $XT$ plots with a 4 mm window, $kf$ spectrum and GOF respectively computed for this window. (h), (i), (j) are $XT$ plot with a 2 mm window, $kf$ spectrum and GOF respectively computed for this window.

## 2.1.2. Harmonic excitation

Quasi-harmonic excitation of guided waves was produced by low-frequency (kHz range) modulation of the high-frequency (1 MHz) push signal described by the expression below:

$$s(t) = \cos^2\left(\frac{\omega}{2}t\right) \times \exp\left[-\left(\frac{\omega(t-t_0)}{4\pi}\right)^4\right], \quad (4)$$

where $\omega$ is the angular frequency of the modulation and $t_0$ a time delay introduced to avoid impulsive loading at time $t = 0$. The geometry of the guided wave excitation in the two-part medium (see Fig. 2a) was the same as for the case of broadband excitation described in the previous section. A typical space-time profile ($XT$ plot) for a 1 kHz excitation is represented in Fig. 2b, and a vertical cross-section (i.e. the distribution of vertical displacement over the propagation distance) for a 6 ms time moment is shown in Fig. 2c.

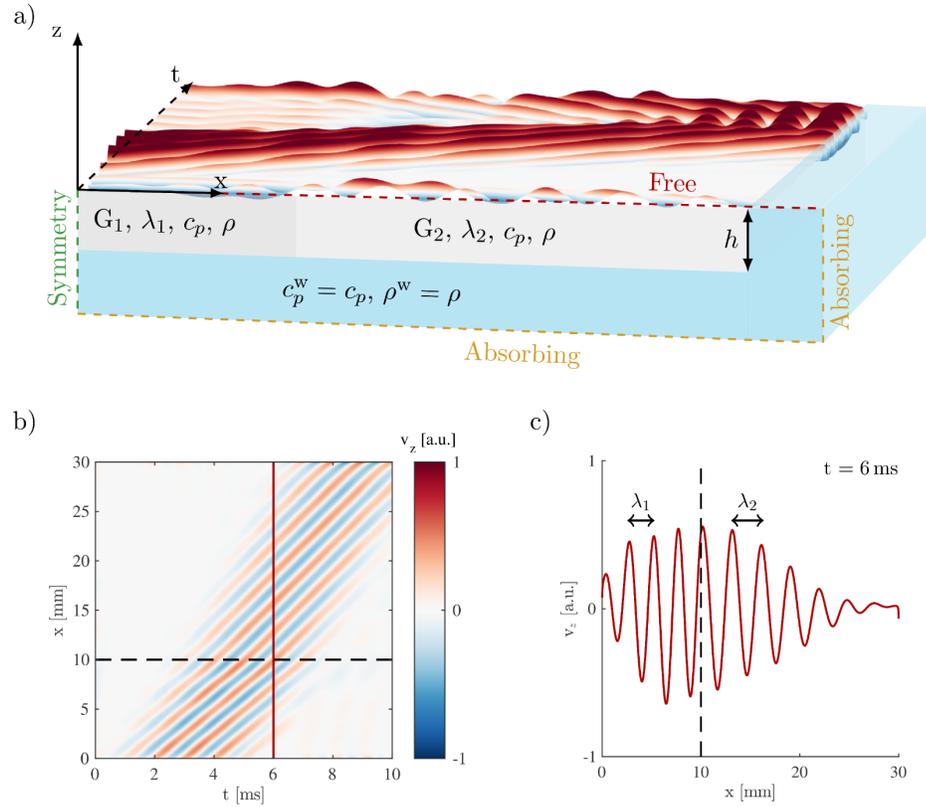

Figure 2. Numerical simulation in OnScale of guided wave propagation in a two-part nearly incompressible linear isotropic material of uniform thickness ($h = 0.5$ mm in this figure) bounded below by water and above by air (mimicking cornea boundary conditions). Quasi-harmonic signals of guided waves are launched with acoustic micro-tapping [10] from air. Shear moduli in parts I and II used in this figure are equal to $G_1 = 30$ kPa and $G_2 = 60$ kPa respectively. The interface between the two parts is at $X = 10$ mm. The push profile is considered Gaussian in space and harmonically modulated in time, as described in Eq. (4). (a) The medium geometry used in the OnScale simulation was the same as for the broadband excitation. (b) Space-time ($XT$) plot of guided wave propagation at the medium surface for a 1 kHz modulation frequency. (c) Vertical vibration velocity as a function of propagation distance at t=6 ms after the excitation. Wavelengths $\lambda_1$ and $\lambda_2$ correspond to regions of part I and part II, respectively.

For quasi-harmonic signals, signal processing described in the previous section is not optimal and unnecessary. Indeed, the wavelength of the propagating wave can be measured directly at every point of the medium using the distance between 2 consecutive signal maxima. As seen in Fig. 2c, the wavelength $\lambda_2$ in the second part of the medium ($X > 10$ mm) is larger than that in the first part according to the difference in mechanical moduli of parts I and II. Wavenumbers in parts I and II can be calculated as reciprocals of wavelengths. The wavenumber computed at every point of the medium at the modulation frequency was then used to determine the shear modulus $G$ using the solution for the $A_0$ mode for this geometry. The dependence of $G$ on the distance from the excitation was then investigated to determine the transition zone between medium parts, i.e. the lateral mechanical resolution.

## 2.2. Two-part phantom

Polyvinyl alcohol (PVA) was used to make a two-component sample with controllable mechanical properties. Optical properties of the phantom were tuned by introducing titanium dioxide micro-particles. The phantom was created using protocols adapted from Ref. [13] and followed the recipe of Ref [1], with a few small modifications. Specifically, two separate concentrated solutions of $TiO_2$ (0.1% wt and 0.075% wt) were prepared in water and dispersed using a magnetic stirrer. 12% wt and 8% wt PVA (146-186 kDa, >99% hydrolyzed, CAS: 9002-89-5, Sigma-Aldrich) was then added to the corresponding $TiO_2$ nanoparticle solutions. The solutions were then covered and stirred on a hot plate at 140°C for approximately two hours to dissolve PVA in solution. Once PVA was fully dissolved, the 12% PVA (.075% $TiO_2$) solution was poured into a circular mold. Pressure was then applied to the surface of the phantom using a glass coverslip to achieve a relatively homogenous, flattened, thin shape and allowed to cool to room temperature (35°C). Once at room temperature, a sharp blade was used to cut the phantom in half and part of the PVA was removed.

Molten 8% PVA (.1% $TiO_2$) was then poured into the mold and a coverslip applied across both parts to create a 2-part boundary with matched thickness of ~0.5 mm. The circular mold had a diameter of 10 cm. Both solutions were degassed using a vacuum chamber prior to pouring into the mold. Once in the mold, the phantom underwent three freeze-thaw cycles (cooled to -20°C then warmed to 35°C) before it was placed in water for up to 12 hours prior to imaging to prevent dehydration. The two-part phantom was then clamped at the edge and placed on top of a water bath (Fig. 3a) to create boundary conditions that result in total internal reflection of elastic waves (air on top, water on bottom, similar to that of the cornea).

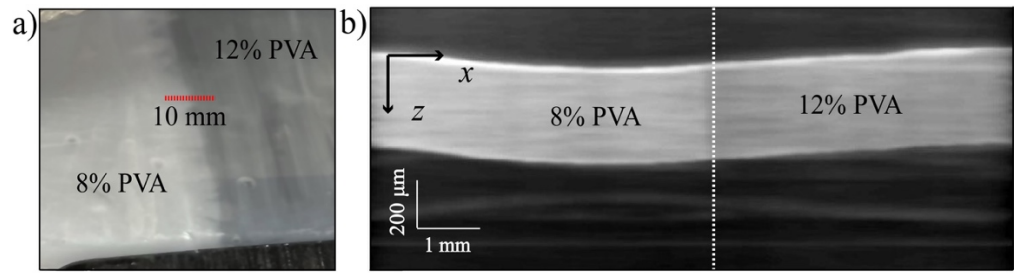

Figure 3. a) The surface of a 2-part PVA phantom with welded boundary, clamped, and suspended on water. The higher concentration of nanoparticles (.1% $TiO_2$) can be seen in the 8% PVA, where the lower concentration of nanoparticles (.075% $TiO_2$) made the 12% PVA more transparent. The red dotted line denotes the approximate OCT B-scan location (not to scale). b) OCT structural image (B-scan) of the 2-part welded phantom. The soft 8% PVA material on the left of the white dotted line appeared brighter due to higher optical scattering; the hard 12% PVA is on the right. The OCT display range was 90-130 dB and a non-uniform aspect ratio was used for visualization.

## 2.3. A$\mu$T-OCE system

A cylindrically focused, resonant (1.05 MHz central frequency) air-coupled AµT transducer was directed at the phantom surface to induce a localized displacement within a thin strip area (approximately 11 mm long and 0.6 mm wide) via reflection-based acoustic radiation force. A temporally short (100 μs), chirped excitation was used to provide approximate pressure confinement in the excitation region. Displacement relaxation resulted in the excitation of quasi-planar broadband pulses of mechanical (both surface and oblique bulk shear) waves of about 0.6 mm duration (i.e. the same as the source width). Detailed information about the AμT transducer and its operating principle can be found in Refs [1], [10], [14]. Due to the wave-reflective interface at the phantom bottom (phantom-water interface), guided waves developed as the elastic wave propagated away from the excitation source. As such, accurate modulus reconstruction required the analysis of guided waves conducted at longer (multiple wavelength) distances from the excitation source.

Elastic waves were tracked in the phantom material using a spectral domain phase-sensitive OCT (PhS-OCT) system operated in M-B mode. A single AμT-OCE scan generated and tracked elastic waves using OCT operating in M-B mode (see details in Ref. [11], [15]). The A-line rate of the system (temporal resolution), determined by the sampling rate of the 1024-pixel line-scan InGaAs array, was 90 kHz, sufficient for alias-free acquisition of induced mechanical waves inside the phantom. The spatial resolution was ~ 10 μm axially and ~ 15 μm laterally. 512 repeated OCT A-scans (referred to as an M-scan) were taken at a set of distances from the AμT focus. Propagating guided waves were tracked with OCT along the surface using a sequence of M-scans conducted at 256 spatial locations over a 10 mm range (dx= 54.7 μm) for every time instant relative to the start of each sequence. The entire spatio-temporal scan with AμT excitations took approximately 1.5 seconds.

The M-B scan dataset was then used to track the propagating wave disturbance based on local vibration velocity. The axial (or vertical) vibration velocity inside the sample at a given location ($\Delta v_z(x, z, t)$) was obtained using the optical phase difference $\Delta\varphi_{opt}(x, z, t)$ between two consecutive A-line scans at each location using the following equation: [16]

$$\Delta v_z(x, z, t) = \frac{\Delta\varphi_{opt}(x, z, t)\bar{\lambda}}{4\pi\bar{n}f_s^{-1}} \quad (5)$$

where $\bar{\lambda}$ was the center wavelength of the broadband light source, $\bar{n}$ was the refractive index of the medium, and $f_s$ was the sampling frequency. The refractive index ($\bar{n}$) was assumed to be 1.3 and the differences due to TiO$_2$ and PVA concentrations were considered negligible.

## 3. Results

### 3.1 Results of numerical simulations

Exploring lateral resolution in OCE for bounded media requires phantoms with predefined mechanical and geometrical properties as well as a tunable OCE system to consider a broad range of excitation parameters, which is not easy to do experimentally. FEM simulations performed on Onscale, however, can closely replicate experimental situations. Such simulations were used to greatly simplify the parametric study of phantom thickness, its mechanical moduli, as well as excitation parameters of the OCE system

### 3.1.1 Broadband excitation

For the boundary conditions of bounded media, multiple guided modes can be generated simultaneously. The appropriate modes should be sorted and analyzed to convert their propagation characteristics into mechanical moduli. Note that the group velocity of propagating guided modes cannot be converted to the material shear modulus [2], [12], and thus different methods must be used. As described above, a common method uses the 2D Fourier Transform of recorded $XT$ plots (Figs.1b, e, h) to compute $kf$ spectra (Figs.1c, f, i). The $kf$ spectra can then be used to identify propagating guided modes. Fitting the computed $kf$ spectrum with a theoretical solution of guided wave propagation can be used to determine the material shear modulus $G$ (see Figs.1d, g, j). The GOF, defining the best estimate of the modulus from the fit, depends on the spatial window size used to compute the $kf$ spectrum. The minimum window size is defined by the threshold where the shear modulus $G$ can still be accurately computed.

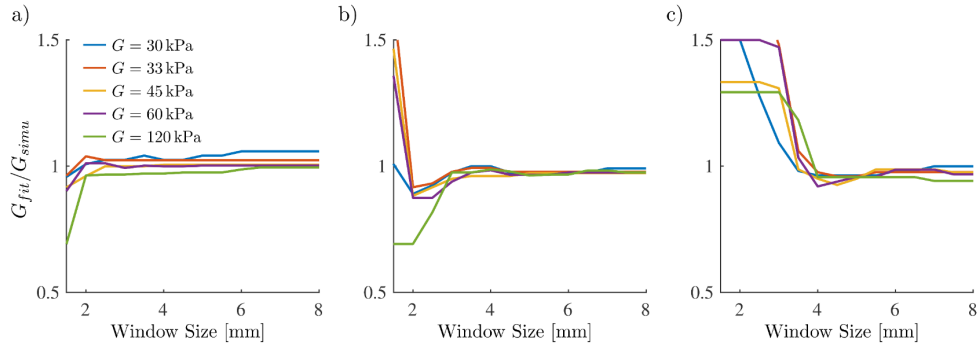

Figure 4. Window-size dependence of the reconstructed value of shear modulus $G$ for different values of $G$ shown in the insert to panel (a) at different layer thicknesses: (a) - $h = 0.25$ mm, (b) - $h = 0.5$ mm, (c) - $h = 1$ mm.

In the FEM simulations, $G_1$ was held constant at 30 kPa and $G_2$ was varied from 30 to 120 kPa. Figure 4 shows the reconstructed shear modulus $G$ as a function of the window size for different values of $G_2$ (curves of different colors) and for different thickness of the bounded material (different panels of Fig. 4). All reconstructions were performed at the center of the second part ($G_2$) (following the $X = 10$ mm interface). It is important to note that when guided waves pass the interface, the spatial duration of the pulse (or mean wavelength) changes because of the change in shear wave speed. For harmonic excitation, this results in a wavelength change (frequency is conserved). Interestingly, the results show that the minimum window size required to accurately reconstruct the elastic modulus does not depend on the shear modulus $G_2$ in the second part of the medium. In other words, the minimum window size for shear modulus reconstruction does not depend on the wavelength of propagating guided waves.

The independence of window size on wavelength may appear to contradict previous results (Ref. [1]) where surface (or Rayleigh) waves were used and the elastographic resolution was determined by the wavelength. There is no contradiction, however, because the ultimate resolution in dynamic OCE is still defined by the mechanical signal wavelength, but guided waves add additional constraints. Importantly, bounded materials require a different mechanical model than surface (Rayleigh) waves that were previously used to reconstruct shear moduli. For the case of a bounded material, sufficient propagation distance is required to form guided modes, which cannot happen over sub-wavelength propagation distances. Correspondingly, Fig. 4 shows that the minimum window size to accurately reconstruct modulus depends on the layer thickness but remains greater than the pulse excitation width.

To explain the 'paradox' presented in Fig. 4, we must first understand acoustic wave generation with A$\mu$T. Impulsive acoustic radiation force applied to the sample surface creates two surface waves (Rayleigh (or SAW) and super-shear evanescent wave (or SEW)[17] waves) and a bulk shear wave propagating at an angle to the surface. In general, the propagation angle of the oblique shear wave depends on the ratio of shear and bulk moduli in the medium [18], but because $G \ll \lambda$ ($\lambda$ is a longitudinal Lamé constant), the directivity pattern for the obliques shear wave does not depend on the shear modulus $G$ and has a maximum at about $\vartheta \approx 26°$ to the sample normal. This explains why the minimum window size is nearly wavelength independent and why it mainly depends on the thickness. Indeed, the propagating oblique shear wave in the layer and its conversion to SEW [17] at each reflection from layer interfaces creates guided waves in the medium. When $\vartheta$ is fixed for all values of $G_2$, the distance at which guided waves are fully formed does not depend on $G_2$. The layer thickness changes the number of reflections per unit length and, thus, changes the minimum distance at which guided waves are formed, or the minimum window size within which guided modes can be identified.

Once the minimum window size needed to identify a guided wave is known, the transition of guided modes through the interface ($X = 10$ mm) can be explored. By translating the specified window through the entire medium length ($0 < X < 20$ mm, see Fig.5a), the reconstructed value of the shear modulus $G$ is mapped (Fig.5b). The modulus change through the transition zone was fit to a combination of two sigmoid functions corresponding to the regions before and after the interface, respectively:

$$C_{R_{fit}} = \frac{a}{1+e^{-b(x-X)}} + c \qquad (6)$$

The derivative of this function can be used to quantify the transition zone length, i.e. determine the lateral spatial resolution of dynamic OCE for bounded media.

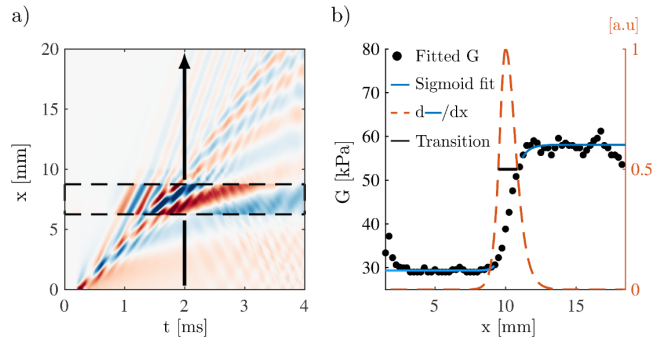

Figure 5. (a) Schematics of moving window, defined in Fig.4, across the interface between two medium parts. (b) Dependence of the reconstructed shear modulus $G$ on the distance from the source (black dots), its fitting with sigmoid function (Eq.6, blue curve) and derivative of the sigmoid function (red dashed curve). Black horizontal bar shows the computed transition zone (or lateral mechanical resolution) at 6dB (full-width half-maximum, FWHM). Elastic moduli of two medium parts used in this figure are $G_1 = 30$ kPa and $G_2 = 60$ kPa; medium thickness is $h = 0.5$ mm.

The relationship between the window size and resolution for materials with different thickness was explored to highlight subtle differences between processing artifact and fundamental limits. Specifically, the 'optimal' window size to reconstruct the modulus can be determined automatically by varying its size and applying it to $XT$ plot processing, as shown in Fig. 6. When the window size was above the optimal one, the transition zone reduced proportional to the window decrease. However, there was a limit where further decrease of the window size did not change the transition zone and, at that, reduced the accuracy of the modulus reconstruction. This also provided a practical method to determine the minimum window size to optimize accuracy without reducing resolution; the optimal window can be taken from the truncation point. The independence of the window size on $G$ is very important because the same window can be used for any gradients of mechanical properties in a bounded medium with constant thickness.

As shown in Fig 6a-c, the optimal window size (or the truncation point in the dependence of the transition zone on window size) strongly depends on layer thickness. Figure 7 shows how the transition zone changed with window size at different layer thicknesses. The relationship between spatial resolution and thickness was obtained for a broad range of $G_2$, and the empirical dependence of spatial resolution on thickness was defined with a linear function:

$$\Delta X = 2.05 \times h + 0.32, \qquad (7)$$

where the transition zone decreased with thickness ($h$) almost linearly over the range of layer thicknesses investigated. The transition zone equaled the spatial resolution in wave-based OCE using guided mechanical waves. Note that in dynamic OCE, OCT is used to track propagating mechanical waves. Thus, the layer thickness can be measured simultaneously and, therefore, the window size can always be properly chosen using the dependences shown in Figs. 6 and 7.

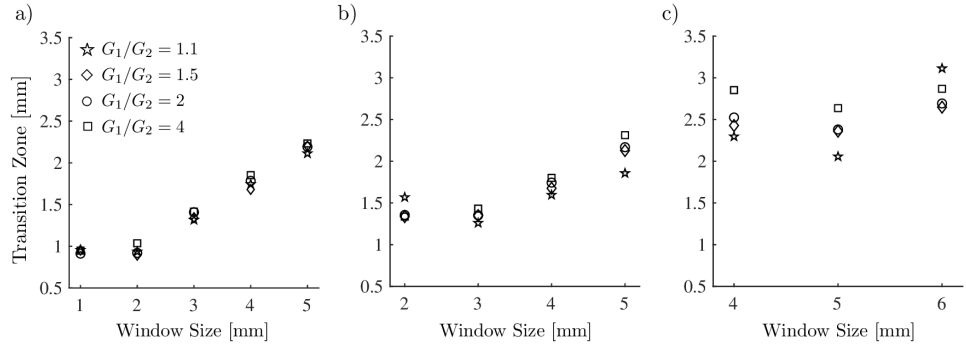

Figure 6. Dependence of the transition zone on window size at different layer thicknesses: (a) - $h = 0.25$ mm, (b) - $h = 0.5$ mm, (c) - $h = 1$ mm. Shear modulus of the first part of the medium used in this figure is $G_1 = 30$ kPa. The ratio of moduli for the two parts is shown in the insert.

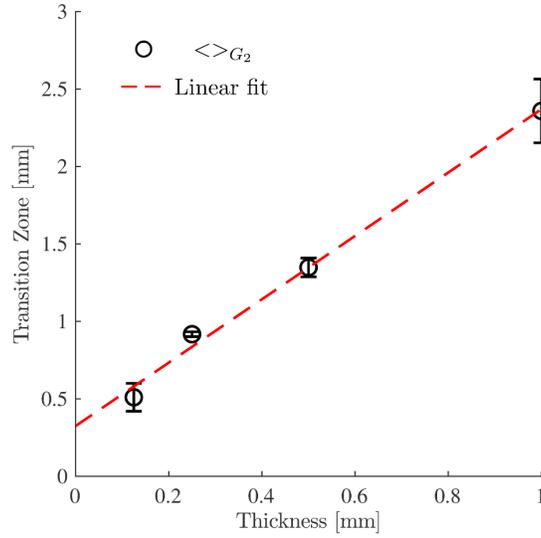

Figure 7. Dependence of the transition zone width on the thickness of a two-part medium. Shear modulus of the first part of the medium used in this figure is $G_1 = 30$ kPa; in the second part shear moduli varied in the range of $G_2 = (33\text{-}120)$ kPa. All depicted values correspond to mean values obtained for different values of $G_2$; the error bars correspond to their variations. The straight line corresponds to fitting the points with a linear function (7)).

### 3.1.2 Harmonic excitation

When a harmonic excitation is employed, the frequency of propagating waves is always fixed, but the wavelength (or wavenumber) changes from point to point depending on the local thickness and mechanical modulus. As we pointed out in the Methods section (see Fig. 2), measuring the local wavenumber at a known sample thickness can be used to reconstruct the mechanical modulus $G$ using the $A_0$ mode dispersion relationship [12]. Using a window to fit the guided wave spectrum in the case of harmonic excitation is meaningless because the wavenumber can be directly determined from the space-time wavefield (see Fig. 2c).

Figure 8 shows the dependence of the lateral mechanical resolution on the carrier frequency of quasi-harmonic excitation for different thicknesses of the two-part medium. First, the computed transition zone, which corresponds to the spatial resolution, depends on the layer thickness in a similar fashion to that illustrated in Fig. 7 for broadband excitation. Second, Figs. 8a, d, g demonstrate that the transition zone is also frequency dependent. The reason for this is in the frequency-dependent wavenumber of the guided waves. The resolution can therefore be improved by increasing the excitation frequency. However, there is a problem in using this strategy. As Figs. 8f, h, i show, increasing the excitation frequency leads to fluctuations in the reconstructed moduli, which are significant in the example shown in Fig. 8i. These fluctuations appear for conditions (frequency and thickness) where both $A_0$ and $S_0$ modes, propagating with different speeds, coexist creating phase disturbances in the wavefield. Determining the proper frequency for OCE studies for quasi-harmonic excitation can be extremely difficult when the medium properties are unknown, especially when it is heterogeneous with varying mechanical properties and thickness.

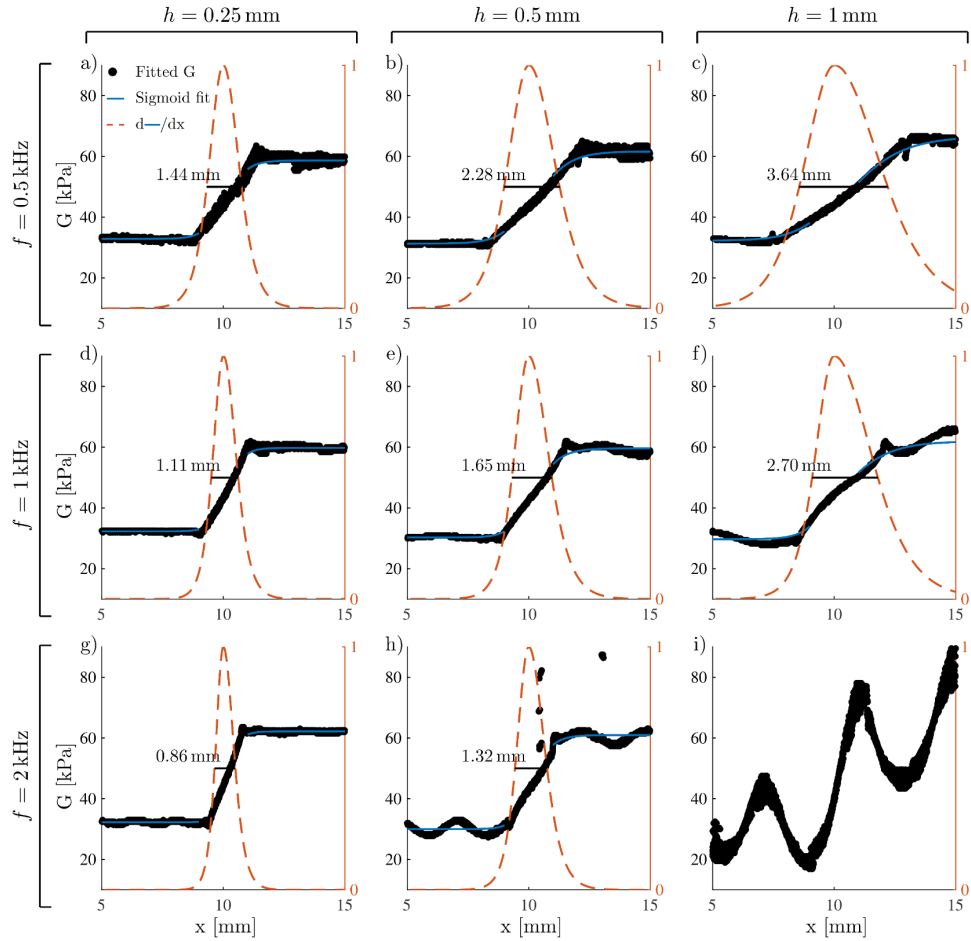

Figure 8. Dependence of the reconstructed shear modulus $G$ on the distance from the source (black dots) in the two-parts medium with shear moduli $G_1 = 30$ kPa and $G_2 = 60$ kPa, respectively, for the case of quasi-harmonic excitation. The excitation frequency and medium thickness are shown on the left and on the top of the panels, respectively. Fits of the transition zone with the sigmoid function (Eq. 6) are plotted with blue curves, and the derivative of the sigmoid function is plotted with the red dashed curve in every panel. The black horizontal bar shows the computed transition zone (or lateral mechanical resolution). The lateral resolution in panel (i) is undetermined due to high fluctuations in reconstructed moduli.

### 3.2 Results of A$\mu$T-OCE experiments

To confirm the findings of numerical simulations, A$\mu$T-OCE experiments were performed in the two-part phantom described in the Methods section. First, wavefields ($XT$ plots) were generated and recorded in each part of the phantom far (~1 cm) from the interface between them. The $XT$ plots in the sections containing only 8% PVA and 12% PVA are represented in Figs. 9a, c respectively. The elastic modulus was reconstructed in both parts using the $kf$ spectra (Figs. 9b, d) computed from the recorded wavefields ($G_1 = (28 \pm 2)$ kPa and $G_2 = (62 \pm 4)$ kPa, respectively). The values and error bars (standard deviation) were computed using fits on 5 different scans with a 10 mm window size. Then, the optimal window size was determined as the minimum range where the reconstruction did not change modulus values. A 4 mm optimal window (Fig. 9e) was found by decreasing the window size in the $XT$ plots. Note that this value was close to that predicted in numerical simulations for a 0.5 mm material. As it was also demonstrated using numerical simulations, the optimal window size was obtained independent of the material modulus, i.e. regardless of the elastic modulus (Fig. 9e).

Finally, propagation of the guided waves was investigated in the direction perpendicular to the interface between the medium parts. The optimal window size (4 mm) was used to compute the medium mechanical properties across the transition between medium parts. Figure 9f demonstrates that an experimentally obtained transition zone is about $\Delta X \cong 1.1$ mm (similar to a $\Delta X \cong 1.3$ mm transition obtained in numerical simulations shown in Fig. 7).

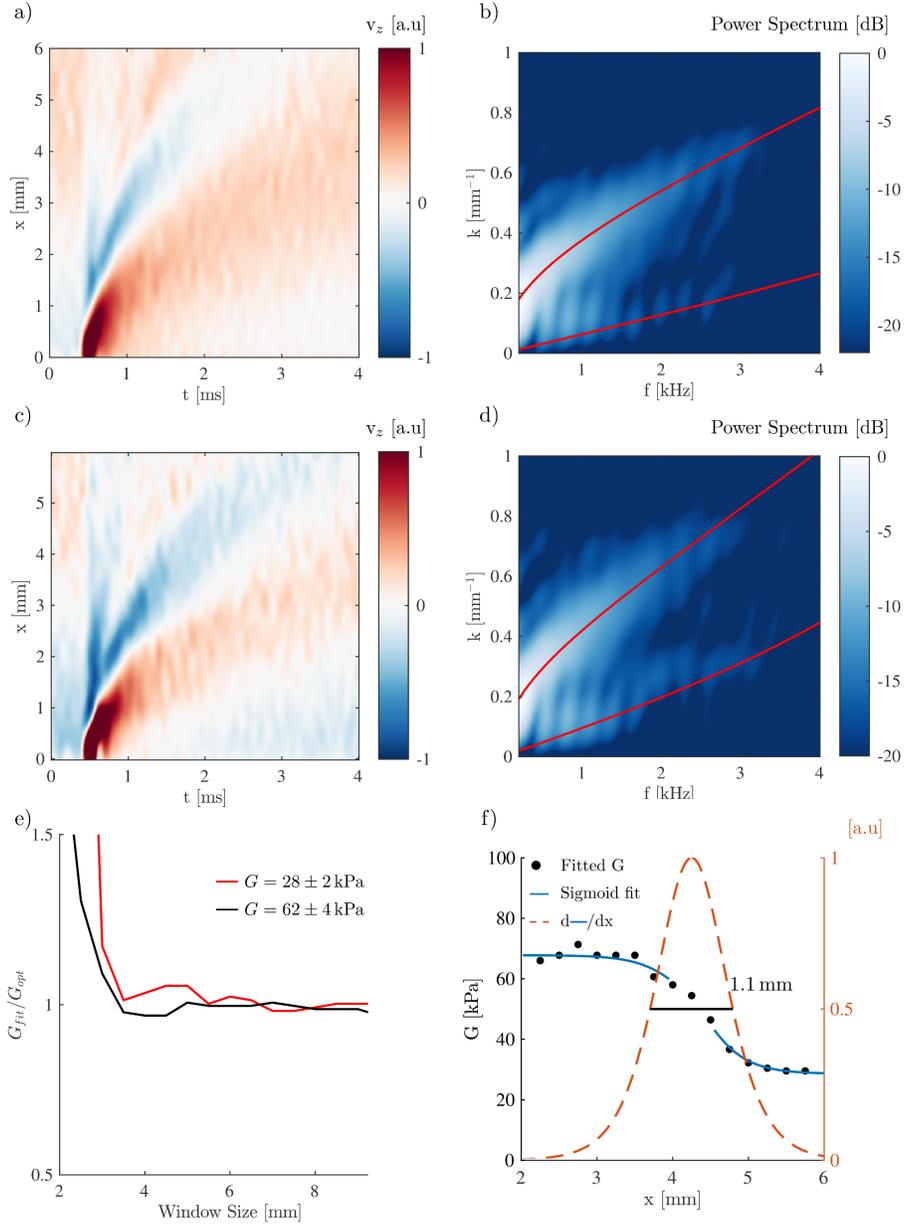

Figure 9. Space-time waveforms (XT plots) recorded experimentally across 6mm in a two-part phantom (see Fig.3) in (a) part I ($G_1 = (28 \pm 2)$ kPa) and (c) part II ($G_1 = (62 \pm 4)$ kPa) and their corresponding $kf$ spectra (b) and (d), respectively. (e) Dependence of the reconstructed shear modulus on window size in each part of the phantom. (f) Dependence of mechanical modulus $G$ on the distance from the A$\mu$T source in the direction from part II to part I across the interface.

### 4. Discussion

In this study, we explored the lateral elastographic resolution of wave-based OCE in bounded soft tissue. Contrary to the broadly accepted opinion that the resolution is fully defined by the capabilities of the imaging system, we demonstrate here that it is not always true. Clearly, geometrical constraints and the appropriate mechanical model play an important role in the ultimate resolution of dynamic OCE.

In Ref. [1] we demonstrated that for Rayleigh waves, the excitation of waves propagating along the interface between materials with different stiffness changes the propagating waves shape. The shape of the propagating wave-front is complex over the region equal to approximately the spatial width of the propagation pulse and, thus, smooths the transition between reconstructed moduli. For this case, the best resolution is not equal to the optical resolution of the OCT imaging system. Practically, it is determined by the propagating elastic wave and can be up to one order of magnitude worse than the optical imaging resolution.

An additional interface, i.e. bounding the tissue, further reduces elastographic resolution in OCE. Guided waves behavior are stimulated in the bounded layer that produce one or multiple dispersive guided modes, with the propagation speed being a function of frequency and thickness. The group velocity of guided modes cannot be as easily converted to the shear modulus as that for the bulk shear wave, and modulus reconstruction from guided wavefields should be performed with care. Guided modes are formed mainly by multiple reflections of the oblique bulk shear wave inside the layer with its conversion to super-shear waves and, therefore, requires a certain distance from the source or stiffness interface for guided modes to form. The propagation speed within the transition zone is variable, and modulus reconstruction within it is inaccurate.

In general, the directivity pattern of the oblique shear wave depends on material properties. For aluminum and brass, for examples, the propagation angle will differ [18]. However, in a nearly incompressible medium, i.e. in soft tissue, the shear modulus is typically five orders of magnitude smaller than the bulk modulus. This leads to the situation in which the angle of bulk shear wave propagation does not depend on the shear modulus [18]. The distance from the source at which the bulk shear wave hits the front surface after its reflection from the bottom of the layer is also relatively independent of the modulus under the ray approximation. Thus, the near field for guided behavior is roughly independent of the wavelength but is dependent on the medium thickness. Whether broadband pulses or harmonic mechanical waves are employed, and regardless of local mechanical moduli, the elastographic resolution in the dynamic OCE of bounded media is mainly determined by the medium thickness.

The relationship between resolution and thickness was demonstrated directly for the case where the thickness was larger than a half of the wavelength (or the spatial pulse width for broadband excitation, see Eq. (7)). The limiting case for decreasing thickness was not explored. In other words, the question remains: how does the transition zone behave when the layer thickness tends to zero? Unfortunately, Fig.7 does not answer this question because the minimum layer thickness we could achieve in numerical simulations was $h = 125\ \mu$m. Further reduction of the thickness led to either numerical dispersion in the solution or further grid refinement (less than 2.5 $\mu$m) keeping all other dimensions in the numerical phantom, which resulted in numerical instabilities. However, additional simulations were performed for a different excitation width (1 mm) for the same layer dimensions in the $(0.125 - 1)$ mm range. The minimum transition zone (or the highest lateral mechanical resolution) we could reach in both series was equivalent to the excitation width. Thus, when the thickness becomes much smaller than the mechanical pulse excitation width, the transition zone reaches its minimum of the excitation width, i.e. the mechanical pulse spatial extent. This connects our previous results obtained with Rayleigh waves with the results of the current study – the best mechanical resolution that can be obtained in dynamic OCE using broadband excitation is on the order of the excitation pulse width.

An important corollary of this work, and the previous exploration using Rayleigh waves, is that the ultimate resolution in dynamic OCE depends on the assumed (or most appropriate) mechanical model and the reconstruction method required to achieve quantitatively accurate contrast. Others [19], for example, showed that the ultimate resolution in a bulk material can be a few times better than half of the wavelength when artifacts induced by the interface are suppressed. For guided waves, however, we do not see a clear way to "unmix" guided modes or analyze them in the near field, accounting for highly variable experimental conditions in real tissue. One possible way to avoid guided waves for modulus reconstruction in bounded media is to use the SEW [17] propagating in the near field along the tissue surface. The wave speed of the SEW is almost twice that of the bulk shear wave. This fast propagation speed suggests that it can be separated from the main guided waves. Unfortunately, the SEW attenuates very quickly, which limits the imaging range or requires multiple pushes over the imaging area.

In this work we considered both broadband and harmonic excitations since both methods are typical for dynamic OCE [11], [20]–[23]. Although harmonic signals generally have better signal-to-noise ratio than broadband pulses, harmonic waves should be treated with care. When a single frequency is generated and material properties are unknown, computation of the local wavenumber can produce serious error due to the presence of multiple guided modes. Here we calculated the local wavenumber assuming that both parts of the phantom propagated only the $A_0$ mode. This, however, is not a given. For the example shown in Fig. 1c, the $S_0$ mode had much higher energy at 4 kHz compared to the $A_0$ mode, whereas at 2 kHz the situation is completely opposite. Thus, if $G_1$ is 4 times higher than $G_2$, the wavenumber calculated for the same frequency in the first part of the medium for the $A_0$ mode will correspond to the wavenumber of the $S_0$ mode in the second part, and the computed modulus will be incorrect. $A_0$ and $S_0$ modes can also coexist, which leads to enormous fluctuations in the reconstructed moduli in both parts of the medium (see Fig. 7i). The same effect will occur when the medium thickness changes. Therefore, harmonic waves at different frequencies should be analyzed in practice to make sure that there are no mode jumps and/or mode competition. Separating reconstruction artifacts from true material heterogeneities may be difficult.

In addition to the local wavenumber calculation, we also explored phase velocity computation, as proposed in Refs [22], [24]. Results for this method can be seen in Supplementary Note 1. Although there appeared to be a certain relationship between medium properties, excitation frequency and thickness, where the spatial resolution appeared to reach the wavelength limit (Suppl. Fig. 1d), it can be extremely difficult to reliably reconstruct modulus using a single frequency in general due to enormous phase instabilities in the guided wavefield (see Suppl. Figs. 1e-i and Suppl. Fig. 2b), except for the case where the probing wavelength is much larger than the layer thickness (Suppl. Fig. 2a). However, using very low frequencies may face a problem of boundary conditions when the medium size is limited, as in cornea, for example. Reducing the frequency will also reduce reconstruction accuracy because low frequencies are highly dispersive. For simple cases where the interface between heterogeneities is sharp and known, high-order guided modes can be filtered out in the $kf$ space, but for unknown borders between heterogeneities or for inclusion small compared to the wavelength, such local filtration may be difficult.

When broadband guided wave signals are used, the 2D Fourier transform can be used to reconstruct wave dispersion spectra and sort guided waves in the $kf$ domain. Fitting a theoretical solution to expected modes in the $kf$ spectrum over a broad frequency range avoids the artifacts associated with single frequency waves. In addition, such processing provides an avenue to ensure that the mechanical model for the medium is chosen correctly, which can be characterized with the goodness of fit (GOF).

As a final tradeoff in determining the resolution in guided wave-based OCE, the processing window should be optimized for accuracy without limiting resolution. Note that when broadband signals are used to reconstruct moduli, quantitative errors are reduced. If the reconstruction window is larger than optimal, this will not change the reconstructed modulus, but will rather smooth the transition between medium heterogeneities. Thus, using a broadband excitation to reconstruct elastic modulus of thin materials in wave-based OCE is advantageous over a single frequency, and can provide mm-scale resolution in many types of human tissue.

## 5. Conclusions

The ultimate elastographic resolution in dynamic OCE has been explored numerically and experimentally for the case of a thin bounded layer, which is typical for OCE applications. There is a broadly accepted opinion that the resolution in mapping of elastic moduli, i.e. elastographic resolution, is equal to that for propagation wave speed, which is in turn determined by the capabilities of the imaging system. We already showed previously that it is not true when waves propagate along the tissue surface (Rayleigh waves), and the accurate reconstruction of mechanical moduli may be corrupted by mode conversions in the region of about a wavelength surrounding the interface or heterogeneity. The situation becomes even more complicated when tissue is layered or bounded, which leads to the generation of dispersive guided modes in the layer. Guided wave propagation makes the phase of harmonic signals unstable and requires a propagation distance before guided modes are fully formed, i.e. there is a near field of the source or interface where wave propagation characteristics are different and, therefore, the reconstruction of mechanical moduli is inaccurate. We have shown here that a minimum transition zone is mainly driven by the medium thickness (being equal to about twice of it) and is nearly independent of the propagation wavelength. Broadband pulse excitation of mechanical waves enables a robust method for modulus reconstruction with the best-achievable resolution using local fitting of wavenumber-frequency spectra with a theoretical model. Processing of harmonic signals for modulus reconstruction should be performed with care to avoid artifacts related to phase instabilities and mode conversions.

*Acknowledgments*

This work was supported, in part, by NIH grants R01-EY026532, R01-EY024158, R01-EB016034, R01-CA170734, R01-AR077560 and R01-HL093140, Life Sci ences Discovery Fund 3292512, the Coulter Translational Research Partnership Program, an unrestricted grant from the Research to Prevent Blindness, Inc., New York, New York, and the Department of Bioengineering at the University of Washington. M. Kirby was supported by an NSF graduate fellowship (No. DGE-1256082). This material was based upon work supported by the National Science Foundation Graduate Research Fellowship Program under Grant No. DGE-1256082.

*Disclosures*

Authors declare no conflict of interests

# Bibliography

[23] F. Zvietcovich and K. V Larin, "Progress in Biomedical Engineering Wave-based optical coherence elastography: The 10-year perspective," 2021.

[24] F. Zvietcovich, J. P. Rolland, J. Yao, P. Meemon, and K. J. Parker, "Comparative study of shear wave-based elastography techniques in optical coherence tomography," *J. Biomed. Opt.*, vol. 22, no. 3, p. 035010, 2017.

# Spatial resolution in optical coherence elastography of bounded media: supplemental document

**Harmonic excitation: computation of the transition zone using local signal phase**

As an alternative to computing the guided wave's local wavenumber as described in Section 3.1.2, local phase difference (or phase velocity) can also be computed. We used numerically simulated wavefields obtained for the same medium parameters described in Fig. 7 of the main paper. We followed the method proposed in Ref. [1], [2] to compute the local phase at every point of the two-part medium. The phase velocity for every medium point was then computed and, finally, its dispersion dependence for the $A_0$ mode was used to calculate the shear modulus $G$. Results are represented in Figure S1 below.

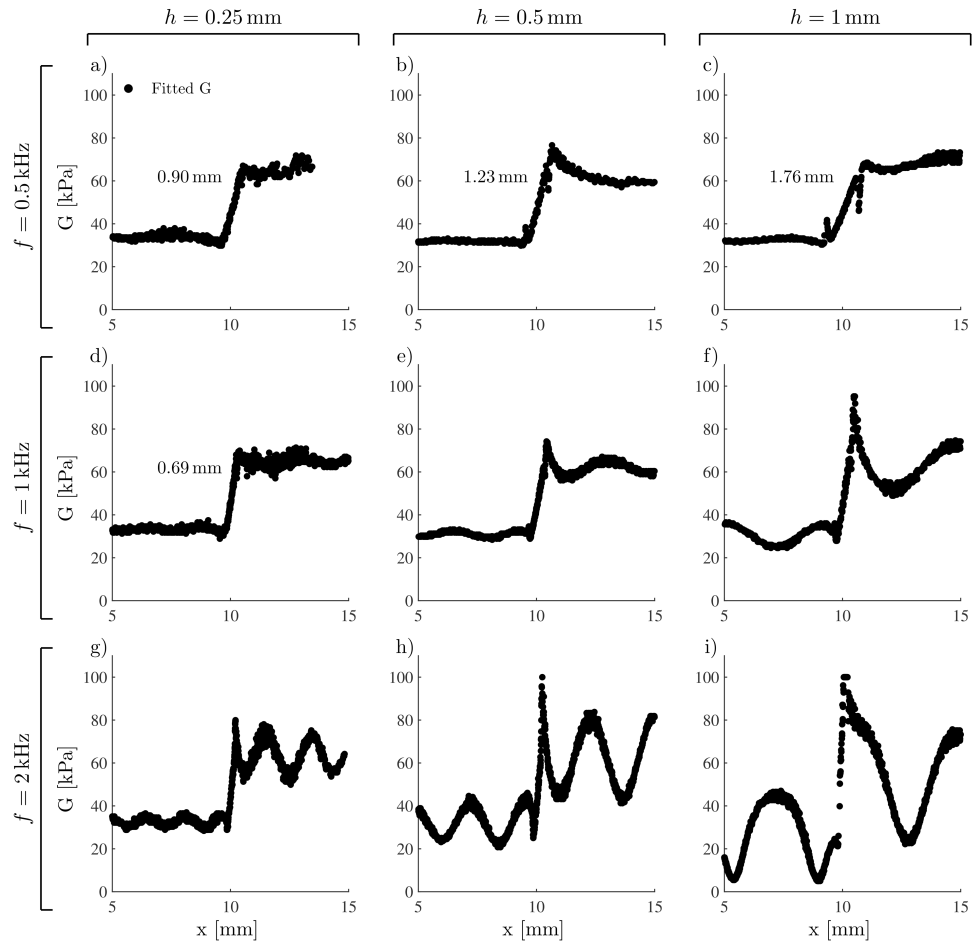

Figure S1. Dependence of the reconstructed shear modulus $G$ on the distance from the source (black dots) in the two-parts medium with shear moduli $G_1 = 30$ kPa and $G_2 = 60$ kPa, respectively, for quasi-harmonic excitation. The excitation frequency and medium thickness are shown on the left and on the top of the panels, respectively. The lateral resolution in panel (e) though panel (i) is undetermined due to high fluctuations in the reconstructed moduli.

As seen in Figure S1, there are significant artifacts in the reconstructed moduli in both parts of the medium due to very large phase instabilities in the wavefield. This result is not surprising because three different wave types (Rayleigh wave, super-shear evanescent wave and bulk shear wave), and their reflections from layer interfaces, mix to form a complex guided wave. Although the frequency of this wave is constant due to the harmonic excitation, mixing of different modes will create a rapidly changing phase and strong amplitude modulation. In terms of guided waves, the fast-changing phase signifies multiple simultaneous guided modes.

Figure S2 shows wavefields for a few different situations. Figure S2a (corresponding to the resolution curve of Fig. S1a) presents a case where only the $A_0$ mode is present. In contrast, Fig. S2b (corresponding to the resolution curve of Fig. S1i) shows a case where both $A_0$ and $S_0$ modes co-exist. The mixing of $A_0$ and $S_0$ modes is clearly seen in the $kf$ spectrum (Figs. S2d, f) compared to the single $A_0$ mode case presented in Figs. S2c, e.

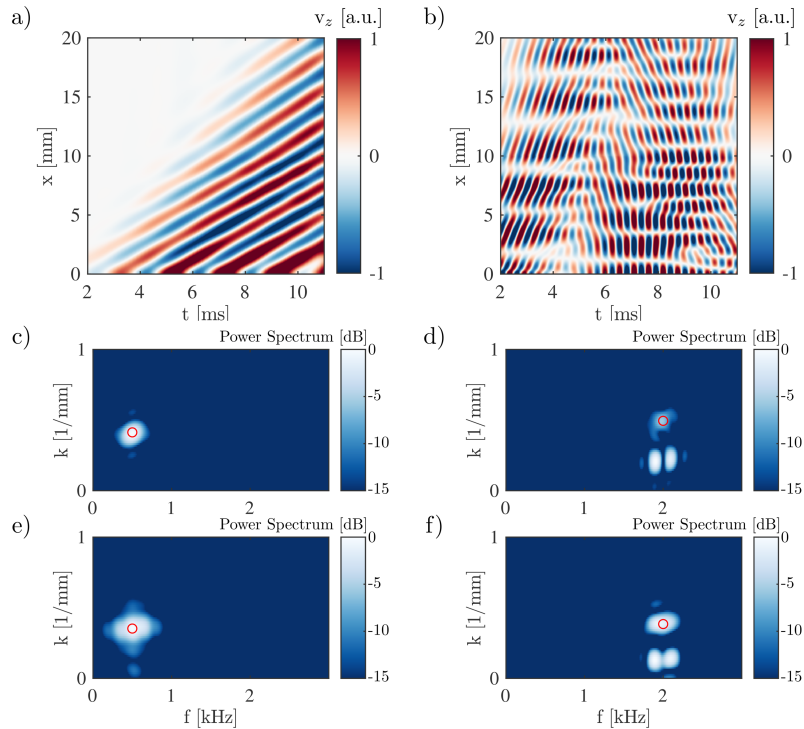

Figure S2. Simulated surface wave fields in the case of quasi-harmonic excitation in a two-part medium with shear moduli $G_1 = 30$ kPa and $G_2 = 60$ kPa and: a) $h = 0.25$ mm and $f = 0.5$ kHz, b) $h = 1$ mm and $f = 2$ kHz. Fourier $kf$ power spectra of the wavefield (a) computed in part I (c) and part II (e) respectively; and Fourier $kf$ power spectra of the wavefield (b) computed in part I (d) and part II (f) respectively.

2222